\documentclass[twoside]{article}

\usepackage{fleqn,espcrc2}
\usepackage{graphicx}
\usepackage{amsmath,amssymb}

\newcommand{{\SD}}{\rm SD}

\newcommand{\vex}{\mbox{\boldmath${\rm x}$}}

\newcommand{\ran}{\rangle}

\newcommand{\xxi}{\mathcal{F}}

\title{Chiral shifts in heavy-light mesons}

\author{A.M.Badalian, Yu.A.Simonov, M.A.Trusov  \address{ITEP, Moscow, Russia}}

\begin{document}

\begin{abstract}
The mass shifts of the $P$-wave $D_s$ and $B_s$ mesons due to
coupling to $DK$ and $BK$  channels are calculated in the coupling
channel model without  fitting parameters. The strong mass shifts
down for $0^+$ and ${1^+}'$ states have been obtained, while
${1^+}''$ and $2^+$ states remain almost in situ. The masses of
$0^+$ and ${1^+}'$ states of $B_s$ mesons have been predicted.
\end{abstract}

\maketitle

After the experimental discovery of the $D_s(2317)$ and
$D_s(2460)$ mesons \cite{1-2}, a necessity to study the chiral
dynamics in heavy-light mesons became quite clear. The masses of
these states proved to be much lower than expected values in
ordinary quark models while their widths were surprisingly small.
The problem was studied in different approaches: in relativistic
quark model calculations \cite{3I-4I}--\cite{6I},  on the lattice
\cite{7I}, in QCD Sum Rules \cite{8I,9I}, in chiral models
\cite{10I,11I-12I} (for reviews see also \cite{13I,14I}). The
masses of $D_s(0^+)$ and $D_s(1^{+'})$ in closed-channel
approximation typically exceed by $\sim$ 140 and 90 MeV their
experimental numbers. The  main theoretical goal  seems for us  to
understand dynamical mechanism responsible for such large mass
shifts of the $0^+$ and $1^{+'}$ levels  and   explain why the
position of other two levels  remains practically unchanged. The
importance of second fact has been underlined by S.Godfrey in
\cite{5I}.

The mass shifts of the $D_s(0^+,1^{+'})$ mesons have already been
considered in a number of papers with the use of unitarized
coupled-channel model \cite{15I}, in nonrelativistic Cornell model
\cite{16I}, in semi-relativistic model with inverse heavy quark
mass expansion \cite{Matsuki}, and in different chiral models
\cite{17I}--\cite{19I}. Here we address again this problem with
the aim to calculate also the mass shifts of the $D_s(1^{+'})$ and
$B_s(0^+,1^{+'})$ states and the widths of the $2^+$ and $1^+$
states,  following the approach developed in \cite{18I}, for which
strong coupling to the S-wave decay channel, containing a
pseudoscalar ($P$) Nambu-Goldstone (NG) meson, is crucially
important. Therefore in this approach principal difference exists
between vector-vector ($VV$) and $VP$ (or $PP$) channels. This
analysis of two-channel system is performed with the use of the
chiral quark-pion Lagrangian which has been derived directly from
the QCD Lagrangian \cite{20I} in the frame of the Field Correlator
Method (FCM) and does not contain fitting parameters, so that the
shift of the $D^*_s(0^+)$ state $\sim$ 140 MeV is only determined
by the conventional decay constant $f_K$.

From the common point of view, due to spin-orbit and tensor
interactions the $P$-wave multiplet of a HL meson is split into
four levels with $J^P =0^+, 1^+_L, 1^+_H, 2^+$ \cite{29I}. Here we
use the notation H(L) for the higher (lower) $1^+$ state because a
priori one cannot say which of them mostly consists of the light
quark  $j=1/2$ contribution. In fact, starting with the Dirac's
$P$-wave levels, one has the states with $j=1/2$ and $j=3/2$. And
$1^+_{L,H}$ eigenstates can be parameterized by introducing the
mixing angle $\phi$:
\begin{equation} \begin{gathered} |1^+_H\ran = \cos \phi |j=\frac12\ran + \sin
\phi|j=\frac32\ran,\\ |1^+_L\ran = -\sin\phi |j=\frac12\ran +
\cos\phi |j=\frac32\ran.
\end{gathered}\label{5-6}\end{equation}
In the heavy-quark (HQ)
limit  the states with $j=\frac32 $ and $j=\frac12$ are not mixed,
but for finite  $m_Q$  they can be mixed even in closed-channel
approximation \cite{10I,29I}.

Taking the meson emission to the lowest order, one obtains the
effective quark-pion Lagrangian in the form
\begin{equation}\Delta L_{FCM} =- \int \psi_i^+ (x) \sigma
|\vex|\gamma_5 \frac{\varphi_a\lambda_a}{f_\pi}
\psi_k(x)d^4x.\label{17}\end{equation} Writing the equation
(\ref{17}) as $\Delta L_{FCM}=- \int V_{if} dt$, one obtains the
operator matrix element for the transition from the light quark
state $i$ (i.e. the initial  state $i$ of a HL meson) to the
continuum state $f$ with the emission of a NG meson
$(\varphi_a\lambda_a)$. Thus we are now able to write the coupled
channel equations, connecting any state of a HL meson  to a decay
channel  which contains another HL meson plus a NG meson.

Consider a complete set of  the states $|f\ran$ in the  decay
channel 2 and the set of unperturbed states $|i\ran$ in channel 1.
One arrives at the nonlinear equation for the shifted mass:
\begin{equation}
m[i]=m^{(0)}[i]-\sum\limits_f \dfrac{|<i|\Hat V|f>|^2}{E_f-m[i]},
\label{21}
\end{equation}
where $m^{(0)}[i]$ is the initial mass, calculated in the
single-channel approximation (assumed to be known), $m[i]$ -- is
the final one, $E_f$ is the energy of the final state, and the
operator $\hat V$ provides the transitions between the channels.
Note, that in our approximation we do not take into account the
final state interaction in the $DK$ system and neglect the
$D$-meson motion. Also, in the w.f. we neglect  possible (very
small) mixing between the $D(1^-_{1/2})$, $D(1^-_{3/2})$ states
and between $D_s(2^+_{3/2})$, $D_s(2^+_{5/2})$ states; physical
$D_s(1^+)$ states can be mixed, though.

In subsequent analysis it is convenient to define the masses with
respect to nearby threshold: $m_{\text{thr}}= m_K+m_D$. So, we
introduce the following notations:
\begin{equation}
E_0=m^{(0)}[D_s]-m_D-m_K,\quad \delta m=m[D_s]-m^{(0)}[D_s],
\end{equation}
\begin{equation}
\Delta = E_0+\delta m=m[D_s]-m_D-m_K,
\end{equation}
where $\Delta$ determines the deviation of the $D_s$ meson mass
from the threshold, and can be complex if a decay to $DK$ pair is
allowed. In what follows we consider unperturbed masses $m_0(J^P)$
of the ($Q\bar q$) levels as given (our results do not change if
we slightly vary their position, in this way the analysis is
actually model-independent).

For further calculations  we should insert the explicit meson w.f.
to the matrix element in (\ref{21}). In our approximation for a HL
meson we consider a light $q$ (or strange $s$) quark with current
(pole) mass $m_{q,s}$ moving in the static field of a heavy
antiquark $\bar Q$, and take its w.f. as a 4-spinor obeying the
Dirac equation with the linear scalar potential and the vector
Coulomb potential with frozen $\alpha_s=\text{const}$:
\begin{equation}
U=\sigma r,\quad V_C=-\dfrac{\beta}{r},\quad
\beta=\dfrac{4}{3}\alpha_s.
\end{equation}

Finally, after long cumbersome calculations which are omitted
here, we arrive at the the following equations to determine meson
masses and widths:
\begin{equation}
\begin{array}{l} D_s(0^+):  \quad
E_0[0^+]-\Delta=\tilde\xxi_0(\Delta),
\\[3mm]
D_s(1^+_L):  \\
 E_0[1^+_L]-\Delta=
\cos^2\phi\cdot\tilde\xxi_0(\Delta)+\sin^2\phi\cdot\tilde\xxi_2(\Delta),
\\[3mm] D_s(1^+_H):
 \\
 E_0[1^+_H]-\Delta=
\sin^2\phi\cdot\tilde\xxi_0(\Delta)+\cos^2\phi\cdot\tilde\xxi_2(\Delta),
\\ \vphantom{\bigg|}
\Gamma[1^+_H]=\sin^2\phi\cdot\tilde\Gamma_0(\Delta)
+\cos^2\phi\cdot\tilde\Gamma_2(\Delta),
\\[3mm] D_s(2^+_{3/2}):
  \\
E_0[2^+_{3/2}]-\Delta= \dfrac{3}{5}\cdot\tilde\xxi_2(\Delta),
\quad \Gamma[2^+_{3/2}]=\dfrac{3}{5}\cdot\tilde\Gamma_2(\Delta),
\end{array}
\label{misha_table_4}
\end{equation}
where $\xxi_{0,2}$ and $\Gamma_{0,2}$ are some universal
functions; definition of those, together with calculation details,
can be found in \cite{BST}.

In our analysis  the 4-component (Dirac) structure of the light
quark w.f. is crucially important. Specifically, the emission of a
NG meson is accompanied with the $\gamma_5$ factor which permutes
higher and lower components of the Dirac bispinors. For the
$j=1/2,P$ -wave and the $j=1/2,S$ -wave states it is exactly the
case that this ``permuted overlap'' of the w.f. is maximal because
the lower component of the first state is similar to the higher
component of the second state and vice-versa, while for the
analogous overlap between $j=3/2,P$ -wave and the $j=1/2,S$ -wave
states the situation is opposite. In the end, it leads to the
functions $\xxi_0$, $\Gamma_0$ being much larger than $\xxi_2$,
$\Gamma_2$ for almost all reasonable values of $\Delta$. Thus the
 large shift of the ${1^+}'$ state with a
concurrent small one for ${1^+}''$ state reveals a natural
explanation (see below).

Now we turn directly to the mass computations. We will take into
account the following pairs of mesons in coupled channels ($i$
refers to first (initial) channel, while $f$ refers to second
(decay) one):


\begin{equation}
\begin{tabular}{cc} $i$ & $f$ \\ \hline \hline $D_s(0^+)$
& $D(0^-)+K(0^-)$ \\ \hline
 $D_s(1^+)$ & $D^*(1^-)+K(0^-)$ \\ \hline $D_s(2^+)$ &
$D^*(1^-)+K(0^-)$ \\ \hline
\end{tabular}
\end{equation}

and analogously for $B$-meson case, with corresponding masses and
threshold values (in MeV):
\begin{equation}
\begin{gathered}
m_{D^+}=1869, \quad m_{D^+}+m_{K^-}=2363, \\
m_{D^{*+}}=2010, \quad m_{D^{*+}}+m_{K^-}=2504,\\
m_{B^+}=5279, \quad m_{B^+}+m_{K^-}=5772, \\
m_{B^*}=5325, \quad m_{B^*}+m_{K^-}=5819. \\
\end{gathered}
\end{equation}

The light quark eigenfunction is calculated numerically via Dirac
equation with the following set of parameters: (the same as in our
previous papers \cite{dirac}):
\begin{equation} \begin{gathered} \sigma=0.18~\text{GeV}^2,\quad \alpha_s=0.39,
\\ m_s=210~\text{MeV},\quad m_q\sim 0~\text{MeV}, \end{gathered}
\end{equation}
The choice of $\sigma$ and $\alpha_s$ is a common one in the frame
of the FCM approach, and the value of the light quark mass really
does not influence here on any physical results because of its
smallness in comparison with the natural mass scale
$\sqrt{\sigma}$. The strange quark mass is taken from \cite{ms},
where it was found from the ratio of experimentally measured decay
constants $f(D_s)/f(D)$; the same value can be obtained by a
renormalization group evolution starting from the conventional
value $m_s(\text{2~GeV}) = 90\pm 15\text{~GeV}$.

\begin{table}
\caption{$D_s(0^+)$-meson mass shift due to the $DK$ decay channel
and $B_s(0^+)$-meson mass shift due to the $BK$ decay channel (all
in MeV)} \label{misha_table_11}
\begin{center}
\begin{tabular}{ccccc} \hline state  & $m^{(0)}$ &
$m^{\text{(theor)}}$ & $m^{\text{(exp)}}$ & $\delta m$  \\
\hline \hline $D_s(0^+)$  & 2475 (30)& 2330(20)& 2317 & -145
\\ \hline $B_s(0^+)$ & 5814(15) & 5709 (15) & {\footnotesize not seen} & -105 \\
\hline
\end{tabular}
\end{center}
\end{table}

\begin{table*}
\caption{The $D_s(1^+)$, $D_s(2^+)$ meson mass shifts and widths
due to the $D^*K$ decay channel for the mixing angle $4^\circ$
(all in MeV)} \label{misha_table_12}
\begin{center}
\begin{tabular}{ccccccc} \hline state & $m^{(0)}$ &
$m^{\text{(theor)}}$ & $m^{\text{(exp)}}$ &
$\Gamma^{\text{(theor)}}_{(D^*K)}$ & $\Gamma^{\text{(exp)}}_{(D^*K)}$ & $\delta m$ \\
\hline \hline $D_s(1^+_H)$ & 2568(15) & 2458(15) & 2460 & $\times$
& $\times$ & -110
\\ \hline $D_s(1^+_L)$ & 2537 & 2535 & 2535(1) & 1.1 & $<1.3$ &
-2  \\
\hline $D_s(2^+_{3/2})$ & 2575 & 2573 & 2573(2) & 0.03 &
{\footnotesize
not seen} & -2\\
\hline
\end{tabular}
\end{center}
\end{table*}

\begin{table*}
\caption{The $B_s(1^+)$, $B_s(2^+)$ meson mass shifts and widths
due to the $B^*K$ decay channel for the mixing angle $4^\circ$
(all in MeV)} \label{misha_table_13}
\begin{center}
\begin{tabular}{ccccccc} \hline state & $m^{(0)}$ &
$m^{\text{(theor)}}$ & $m^{\text{(exp)}}$ &
$\Gamma^{\text{(theor)}}_{(B^*K)}$ & $\Gamma^{\text{(exp)}}_{(B^*K)}$ & $\delta m$ \\
\hline \hline $B_s(1^+_H)$ & 5835(15) & 5727(15) & {\footnotesize
not seen} & $\times$ & $\times$ & -108
\\  \hline $B_s(1^+_L)$ & 5830 & 5828 & 5829 (1)& 0.8 & $<2.3$ &
-2
\\
\hline $B_s(2^+_{3/2})$ & 5840 & 5838 & 5839(1) & $<10^{-3}$ &
{\footnotesize not seen} & -2 \\
\hline
\end{tabular}
\end{center}
\end{table*}

The ultimate results of our calculations are presented in Tables
\ref{misha_table_11}--\ref{misha_table_13}. A priori one cannot
say whether the $|j=\frac12\ran$ and $|j=\frac32\ran$ states are
mixed or not. If there is no mixing at all, in this case the width
$\Gamma(D_{s1}(2536))= 0.3$ MeV is obtained in \cite{35}, while
the experimental limit is $\Gamma<2.3$ MeV \cite{26I} and recently
in \cite{36} the width $\Gamma=1.0\pm 0.17$ MeV has been measured.
Therefore small mixing is not excluded and here we take the mixing
angle $\phi$ slightly deviated from  $\phi=0^{\circ}$ ( no mixing
case). Then we define those angles $\phi$ which are compatible
with experimental data for the  masses and widths of both $1^+$
states.

The large value $\cos^2\phi$ for the $1_{H}^+(j=1/2)$ state
provides large mass shift ($\sim 100$ MeV) of this level and at
the same time does not produce the  mass shift of the $1^{+}_L$
level, which is almost pure $j=\frac32$ state. We would like to
stress here that the mass shifts weakly differ for $D_s$ and
$B_s$, or, in other words, weakly depend on the   heavy quark
mass.

Thus we have obtained the shifted masses $M(B_s,0^+)=5710(15)$ MeV
and $M(B_s,1^{+'})=5730(15)$ MeV, which are in agreement with the
predictions in \cite{14I} and of S.Narison \cite{9I} and by $\sim
100$ MeV lower than in \cite{3I-4I},\cite{10I}. The masses of the
$2^+$ and $1^+$ states  precisely agree with experiment.

\section*{Acknowledgments}
The authors would like to acknowledge support from the President
Grant No. 4961.2008.2 for scientific schools. One of the authors
(M.A.T.) acknowledges partial support from the President Grant No.
MK-2130.2008.2 and the RFBR for partial support via Grant No.
06-02-17120.


\begin{thebibliography}{99}

\bibitem{1-2} B. Aubert et al. (Babar Collab.), Phys. Rev. Lett. {90} (2003) 242001;
D. Besson et al. (CLEO Collab), Phys. Rev. {D 68} (2003) 032002;
P. Krokovny et al. (BelleCollab), Phys. Rev. Lett. {91} (2003)
262002.
\bibitem{3I-4I} S. Godfrey, N. Isgur, Phys. Rev. {D 32} (1985) 189;
S. Godfrey, R. Kokoski, Phys. Rev. {D 43} (1991) 1679;
D. Ebert, V. O. Galkin, R. N. Faustov, Phys. Rev. {D 57} (1998)
5663 [Erratum: ibid. {D 59} (1999) 019902].
\bibitem{5I} S. Godfrey, Phys. Rev. {D 72} (2005) 054029.
\bibitem{6I} Yu. S. Kalashnikova, A. V. Nefediev, Yu. A. Simonov, Phys.
Rev. {D 64} (2001) 014037; Yu. S. Kalashnikova, A. V. Nefediev,
Phys. Lett. {B 492} (2000) 91.
\bibitem{7I} R. Lewis, R. M. Woloshyn, Phys. Rev. {D 62} (2000) 114507;
G. S. Bali, Phys. Rev. {D 68} (2003) 0715001; A. Dougall et. al ,
Phys. Lett. {B 569} (2003) 4.
\bibitem{8I} Y. B. Dai,  C. S. Huang,
C. Liu, S. L. Zhu, Phys. Rev. {D 68} (2003) 114011.
\bibitem{9I} S. Narison, Phys. Lett. {B 605} (2005) 319.
\bibitem{10I} M. Di Pierro, E. J. Eichten, Phys. Rev. {D 64} (2001) 114004.
\bibitem{11I-12I} W. A. Bardeen, E. J. Eichten, C. T. Hill, Phys. Rev.
{D 68} (2003) 054024; K. D. Chao, Phys. Lett. {B 599} (2004) 43.
\bibitem{13I} E. S. Swanson, Phys. Rept. {429} (2006) 243.
\bibitem{14I} P. Colangelo, F. De Fazio, R. Ferrandes, Mod. Phys. Lett.
{A 19} (2004) 2083; P. Colangelo, F. De Fazio, Phys. Lett. {B 570}
(2003) 180, hep-ph/0609072.
\bibitem{15I} E. van Beveren, G. Rupp, Phys. Rev. Lett. {91} (2003) 012003;
Mod. Phys. Lett. {A 19} (2004) 1949.
\bibitem{16I} D. S. Hwang, D. W. Kim, Phys. Lett. {B 601} (2004) 137.
\bibitem{Matsuki} T. Matsuki and T. Morii, Phys. Rev. {D 56} (1997) 5646.
\bibitem{17I} F. L. Wang,  X. L.Chen, D. H. Lu, S. L. Zhu, W. Z. Deng,
 hep-ph/0604090; Y. B. Dai, S. L. Zhu, Y. B. Zuo, hep-ph/0610327.
\bibitem{18I} Yu. A. Simonov, J. A. Tjon, Phys. Rev. {D 70} (2004) 114013.
\bibitem{19I} J. Vijande, A. Valcarse, F. Fernandez, Phys. Rev D 77 (2008) 017501.

\bibitem{20I} Yu. A. Simonov, Phys. Rev. {D 65} (2002) 094018.



\bibitem{29I} R. N. Cahn, J. D. Jackson, Phys. Rev. D 68 (2003) 037502 .

\bibitem{BST} A. M. Badalian, Yu. A. Simonov, and M. A. Trusov, Phys.
Rev. {D 77} (2008) 074017.





\bibitem{dirac} Yu. A. Simonov and M. A. Trusov, hep-ph/0506058,
hep-ph/0607075.

\bibitem{ms}  A. M. Badalian, B. L .G. Bakker,  arXiv:hep-ph/0702229.



\bibitem{35} A. F. Falk, T. Mehen, Phys. Rev. {D 53} (1996) 231;
P. L. Cho, M. B. Wise, Phys. Rev. {D 49} (1994) 6228.

\bibitem{26I} Particle Data Group, J. of Phys. {G 32} (2006) 1.

\bibitem{36} A. Zghiche ( for BaBar Collaboration), arXiv:0710.0314.

\end{thebibliography}
\end{document}